# Evidences of the earliest Supernova observation in the Nebra disk

## R.G. Pizzone

The Nebra disk, found in Northern Germany, is one of the first representation of astronomical objects we have. It is represented in figure 1 and the most striking features are, without any doubt, the Pleiades cluster, which shows up in the middle of the disk. A "solar disk" and a waning moon also appear in a starry background. It was shown [Moeller 2004] that some other arcs were added later to the original figure. Presumably, from $^{14}$C dating, the disk was buried around 1600-2000 BC, being made generations before.

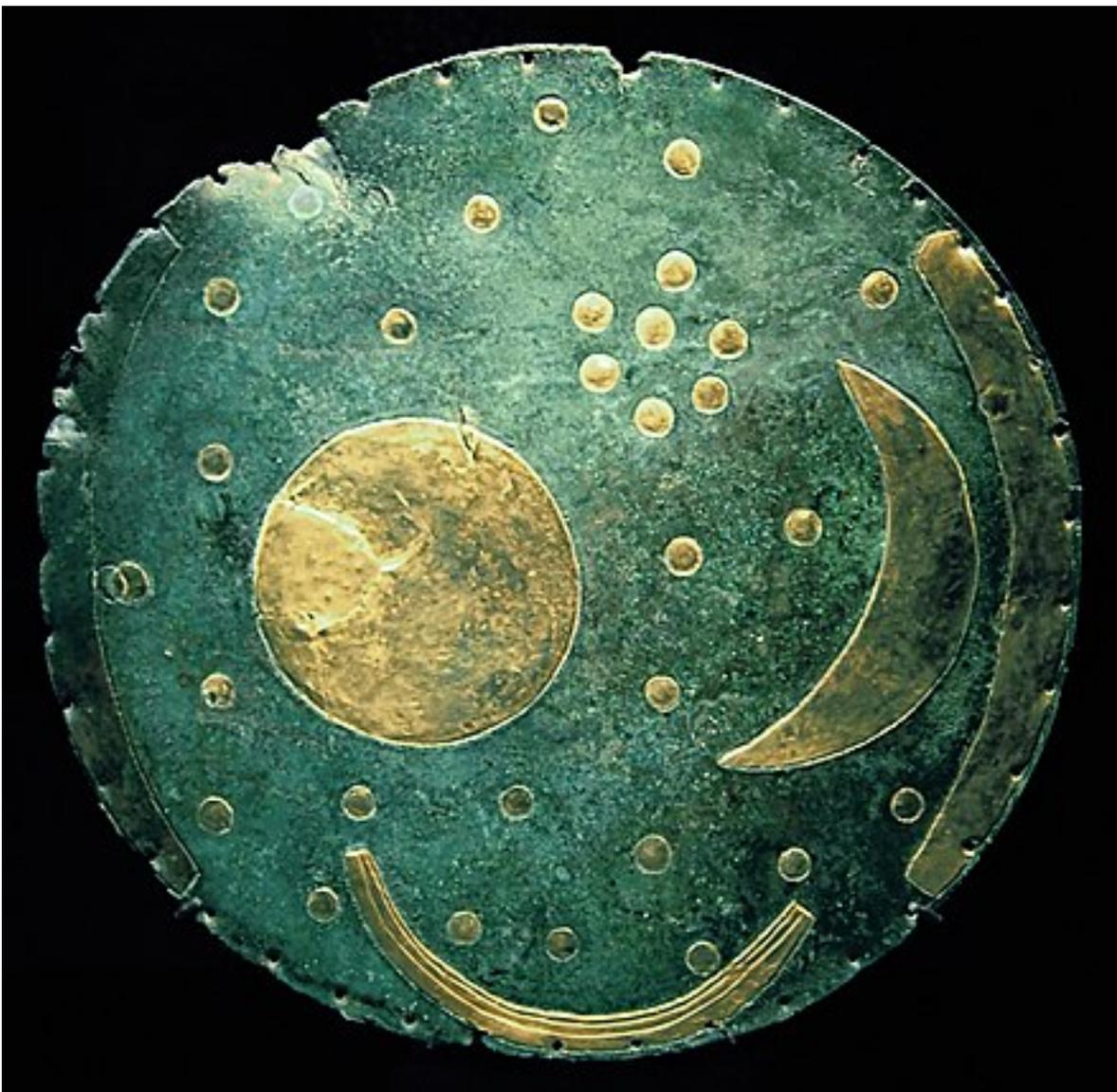

**Figure 1.** The Nebra disk as it appears now (on display in Halle Museum of Prehistory). The Pleiades cluster shows up in the top part, amid the golden disk and the waning Moon.

Several studies have been performed to understand the eventual astronomical significance of the object e.g. in Pásztor 2015 and references therein. Nevertheless, expecially in the first version on the disk with just the starry background, the golden disk and the moon as well as the Pleiades an astronomical correlation with any particular phenomenon is hardly found.

We reinterpret the picture with emphasis on the two extremely bright celestial bodies shown in the picture. There is clear indication that the two celestial objects drawn are very bright. One of the objects is either the Sun and second object is a crescent Moon, relatively close to the first. They cannot be Sun and Moon since, with such proximity to the Sun, the Moon would be in a partial phase around the new and hence not very bright. We investigate the possibility that the observed object is a star as even in other prehistoric drawings from European caves, stars are never shown as large disks. Moreover starts are plotted (e.g. the Pleiades or background stars) as smaller disks. Therefore it seems the picture represented in the Nebra disk deals with a night starry background, in the sky region of the Pleiades. A very luminous object is present close to the seven sisters and the crescent Moon.

Our idea is that the ancient people who made the disk did have an astronomical phenomenon to describe and, by portraying it, worship for the ages to come. The singular phenomenon is a SuperNova explosion (symbolized by the large golden disk) in the region of the sky which is closer to the Pleiades, i.e. the Taurus/Auriga asterism. That area of the sky sits in a very active region for Stellar formation of our Galaxy and therefore in an area where several Supernovae events have occurred historically. The best known is the 1054 Supernova which formed the Crab nebula. We underwent an extensive research of any supernova remnant (SNR) which is in that area and with an age which matches the disk one. In table 1 the results of this search are reported, noting also the SNR age and distance while their relative position with respect to the Pleiades is reported in figure 2.

| SuperNova Remnant | Age (years) | Distance (parsec) | Reference |
|---|---|---|---|
| SH2 224 | 13000 -24000 | 4500 | Burrows &Guo 1994 |
| SH2 221 (or HB9) | 4000-6000 | 400-800 | Lehay&Tian 2007 |
| Simeis 147 | 40000 | 1500 | Romani et al 2007 |
| Crab Nebula | 964 | 2000 | Kaplan 2008 |
| IC443 | 3000-30000 | 1500 | Chevalier 1999 |

**Table 1:** Supernova remnants which lay in the Auriga/Taurus area. Age and distance are also reported.

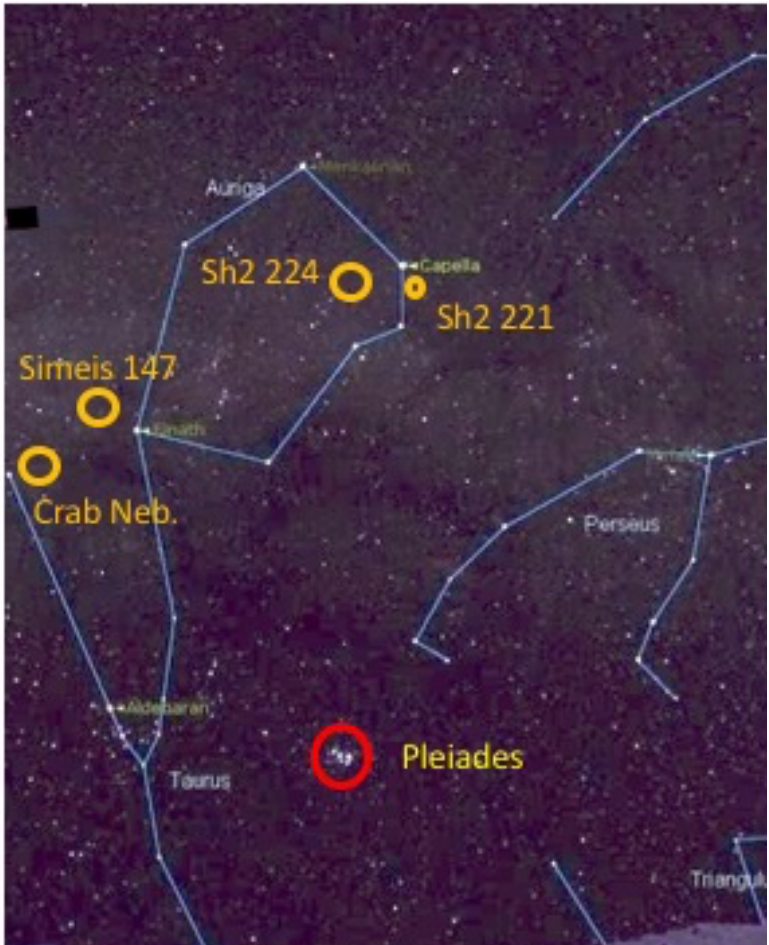

**Figure 2:** Position in the Taurus – Auriga region of the SNR in table 1. IC443 is not reported, being to the left (in Gemini).

It turns out that SH2-221, a supernova remnant which was recently object of heavy study (Lehay & Tian 2007), has an age which might be compatible with the Nebra Disk fabrication. Being the explosion relatively close to the Sun (400-800 pc) and assuming an absolute magnitude for a Typical type II supernova (which was at the origin of SH2-221 and the pulsar slightly off its center) $M_V = -17$ (at peak) we can easily calculate the apparent magnitude of the Supernova explosion at its peak. This yelds a brutal $-9 < m_V < -7.5$ which is relatively close to the Moon (-12.7). Being this a celestial phenomenon of such magnitude it seems reasonable that it was taken as a possible herald of important events (as until recent years in several cultures for the comets). Therefore it was portrayed on the Nebra Disk together with the most interesting objects nearby (the Moon and the Pleiades).

With time as the memory of the Supernova event slowly cancelled, generation after generation, maybe a new solar symbolism was found for the old object and solar arcs were drawn as possible hints to astronomical phenomena like solstices.

Therefore the Nebra disk may be the first representation of a Supernova explosion ever made by mankind. Of course it cannot be demonstrated although it might be reasonable. This seems in agreement with a similar explanation of stone carvings, made in the same period in the Burzahama region (India) as reported in reference Hrishikesh J et al. 2007.

**References**
Burrows & Guo Ap. J. 421 L19-22 1994
Chevalier R., Ap. J. 511, 798, 1999
Hrishitesh J et al., Indian Journal of History of Science 42.1 (2007) 83
Kaplan DL et al., Ap. J. 677, 1201 (2008)
Lehay &Tian Astronomy and Astrophysics 461 1013 (2007)
Moeller H, National Geografic 76-8 (2004)
Pasztor E. "Nebra disk" Handbook of Archeoastronomy and Ethnoastronomy, Springer, 1349 (2015)
Romani et al., Ap. J., 654 487-493 2007